\documentclass[aps,prd,nofootinbib,twocolumn,showkeys,floatfix]{revtex4-1}

\usepackage{graphicx}
\usepackage{amsmath,amssymb}
\usepackage{datetime}
\usepackage{helvet}
\usepackage{pstricks}
\usepackage{graphicx,color}
\usepackage{amssymb,amsmath}  
\usepackage[colorlinks=true,allcolors=SBred, pdfborder={0 0 0}]{hyperref}

\usepackage{booktabs}
\renewcommand{\toprule}{\specialrule{1.5pt}{0em}{0em}}
\renewcommand{\bottomrule}{\specialrule{1.5pt}{0em}{0em}}

\definecolor{SBred}{rgb}{0.6471, 0.1098, 0.1882}
\def\gsim{\raise0.3ex\hbox{$\;>$\kern-0.75em\raise-1.1ex\hbox{$\sim\;$}}}
\def\lsim{\raise0.3ex\hbox{$\;<$\kern-0.75em\raise-1.1ex\hbox{$\sim\;$}}}
\newcommand{\order}[1]{\mathcal{O}(#1)}
\def\vev#1{\left\langle #1\right\rangle}
\newcommand{\eq}[1]{eq.~(\ref{eq:#1})}

\newcommand{\app}[1]{app.~(\ref{app:#1})}

\newcommand{\tab}[1]{table~\ref{tab:#1}}

\newcommand{\secs}[1]{sec.~\ref{sec:#1}}
\newcommand{\nn}{\nonumber}
\def\one{\ensuremath{\mathbf{1}}}

\def\three{\ensuremath{\mathbf{3}}}
\def\threeS{\ensuremath{\mathbf{3^*}}}	
\def\EW{SU(2)_L \otimes U(1)_Y}
\def\TrTrOne{SU(3)_c \otimes SU(3)_L \otimes U(1)_\mathcal{X}}
\def\TTOEW{SU(3)_L \otimes U(1)_\mathcal{X}}
\def\hc{\mathrm{h.c.}}
\def\z2{\mathbf{Z}_2}
\def\tr{\top}
\newcommand{\gev}{\,\mathrm{GeV}}
\newcommand{\tev}{\,\mathrm{TeV}}

\begin{document}

\title{The LHC diphoton resonance from gauge symmetry}
\author{Sofiane M. Boucenna}
\email{boucenna@lnf.infn.it}
\affiliation{INFN, Laboratori Nazionali di Frascati, C.P. 13, 100044 Frascati, Italy.}
\author{Stefano Morisi}
\email{stefano.morisi@gmail.com}
\affiliation{Dipartimento di Fisica, Universit\'a di Napoli ``Federico II'' and INFN, Sezione di Napoli, Complesso Univ. Monte S. Angelo, Via Cintia, I-80126 Napoli, Italy}
\author{Avelino Vicente}
\email{avelino.vicente@ific.uv.es}
\affiliation{Instituto de F\'{\i}sica Corpuscular (CSIC-Universitat de Val\`{e}ncia), Apdo. 22085, E-46071 Valencia, Spain.}

\begin{abstract}
\noindent
Motivated by what is possibly the first sign of new physics seen at the LHC, the diphoton excess at $750$ GeV in ATLAS and CMS,
we present a model that provides naturally the necessary ingredients to explain the resonance.
The simplest phenomenological explanation for the diphoton excess
requires a new scalar state, $X(750)$, as well as additional
vector-like (VL) fermions introduced in an ad-hoc way in order to enhance its decays
into a pair of photons and/or increase its production cross-section. We
show that the necessary
VL quarks and their couplings  can emerge naturally from
a complete framework based on the $\TTOEW$ gauge symmetry.
\end{abstract}

\maketitle

\section{Introduction and motivations}
\label{sec:intro}
The great expectations to find New Physics (NP) at the LHC may have materialized with the observation 
of a diphoton excess at $\sim 4\sigma$ at $\sim 750$ GeV by the ATLAS and CMS collaborations~\cite{ATLAStalk,CMStalk}.
This signal, if confirmed by further data, would
be the first clear and direct LHC indication of physics beyond the
Standard Model (SM), a framework that was anyway expected to be
incomplete due to its
failure to account for non-zero neutrino masses, dark matter, and 
the matter-antimatter asymmetry of the Universe.

\bigskip

Given the low statistical significance of this hint (locally $3.6
\sigma$ and $2.6 \sigma$ in ATLAS and CMS, respectively), it is
definitely too soon to claim the end of the SM. Nevertheless, it is
tempting to speculate about the possible origin of such an excess. In
fact, the announcement of this hint has triggered an intense activity
translated into many recent papers analyzing the diphoton excess and
proposing various physical explanations as to its origin~\cite{DiChiara:2015vdm,Franceschini:2015kwy,Pilaftsis:2015ycr,Buttazzo:2015txu,Knapen:2015dap, Angelescu:2015uiz,Backovic:2015fnp,Mambrini:2015wyu,Harigaya:2015ezk,Bai:2015nbs,Falkowski:2015swt,Csaki:2015vek,
Chakrabortty:2015hff,Bian:2015kjt,Curtin:2015jcv,Fichet:2015vvy,Chao:2015ttq,Demidov:2015zqn,Martinez:2015kmn,No:2015bsn,Ahmed:2015uqt,Cox:2015ckc,Kobakhidze:2015ldh,Cao:2015pto,Dutta:2015wqh,Ellis:2015oso,Petersson:2015mkr,Low:2015qep,McDermott:2015sck,Higaki:2015jag,Matsuzaki:2015che,Aloni:2015mxa,Bellazzini:2015nxw,Gabrielli:2015dhk,Benbrik:2015fyz,Kim:2015ron,Alves:2015jgx,Carpenter:2015ucu,Bernon:2015abk,Nakai:2015ptz,Molinaro:2015cwg,Huang:2015evq,Han:2015qqj,Cao:2015twy,Ding:2015rxx,Barducci:2015gtd,Chala:2015cev,Bauer:2015boy,Cline:2015msi,Cho:2015nxy,Berthier:2015vbb,Kim:2015ksf,Bi:2015uqd,Dhuria:2015ufo,Liao:2015tow,Wang:2015kuj,Antipin:2015kgh,Heckman:2015kqk,Chang:2015sdy,Chao:2015nsm,Bardhan:2015hcr}. Generally speaking, the simplest new physics interpretation of the diphoton
excess is through the radiative decay of a new spin-0 state produced resonantly at the LHC.

\bigskip

In what follows we will assume that indeed NP is at work here and the resonance, 
to which we refer as $X(750)$, is genuine.
In contrast with most
explanations to the diphoton excess proposed so far, which introduce
new {ad-hoc} states to enhance the diphoton rate or increase the
$X(750)$ production cross-section,
we will contemplate the possibility that this particle, as well as the
necessary ingredients to get the required diphoton signal, are the
result of some gauge extension of the SM.\\

\bigskip

Perhaps, the simplest phenomenological extension of the SM that can
account for $X(750)$ is the addition of a {real} scalar singlet, that we
denote as $X$, and a vector-like (VL) quark, $\mathcal{Q}$. The combination of
these two elements allows us to write a phenomenological Lagragian,
\begin{equation}
\label{eq:lagPheno}
 - \mathcal{L}_\text{pheno} =\frac{1}{2}
 M_X X^2 + M_Q \,\overline{\mathcal{Q}} \mathcal{Q} +\lambda\, X \overline{\mathcal{Q}} \mathcal{Q} \,,
\end{equation}
which effectively generates the interactions with gluons and photons,
$c_s\, X\, G_{\mu\nu}G^{\mu\nu} + c_e X\, F_{\mu\nu} F^{\mu\nu}$, with
$c_a \propto \frac{\lambda \alpha_a^2}{M_{Q}}$ for $a=s,e$ { and $\alpha_{s(e)}$ is the strong (electromagnetic) coupling strength}.

It is our goal here to generate such effective interactions from a
gauge extension of the SM.
A simple embedding of $\EW$ into a larger group is provided by
$\TTOEW$. The group structure forces the introduction of new colored
fermions to complete the $SU(3)_L$ multiplets. These new quarks are
$\EW$ singlets after the breaking of $\TTOEW$, and offer the
attractive possibility to account for \eq{lagPheno} from the gauge
symmetry. Indeed{, if we take for instance a singlet right-handed quark field $\mathcal{Q}_R$}, a ``weak'' quark multiplet, 
and a scalar in the fundamental representations of $SU(3)_L$:
\begin{equation}
Q_L = \left( \begin{array}{c}
q \\
\mathcal{Q} \end{array} \right)_L \,,\,
\Phi = \left( \begin{array}{c}
H \\
X \end{array} \right) \,, \quad 
\end{equation}
which after the breaking $\TTOEW \to \EW$ gives the quarks and Higgs $SU(2)_L$ doublets, $q_L$ and $H$, as well as the iso-singlets  $\mathcal{Q}$ and $X$.  We see that   { the gauge-invariant} coupling 
$\Phi\, \overline{Q_L}\,\mathcal{Q}_R$ automatically generates at low energies the coupling $ X
\overline{\mathcal{Q}} \mathcal{Q}$ as required in \eq{lagPheno}.\\

In addition to this attractive feature, models based on $\TrTrOne$ 
gauge symmetry (3-3-1 for short)~\cite{Singer:1980sw,valle:1983dk,Pisano:1991ee,Frampton:1992wt} constitute a minimal extension of the SM that could
explain the number of generations,
and provide mechanisms to generate small neutrino masses either
radiatively~\cite{Boucenna:2014ela,Okada:2015bxa} or at
tree-level with new testable flavor
predictions~\cite{Boucenna:2015zwa} and gauge bosons physics lying at the TeV scale. This can
also be related to gauge coupling
unification~\cite{Boucenna:2014dia} {and interesting D-brane constructions~\cite{Addazi:2015eca}.}

\bigskip

The paper is organized as follows. In the next Section we present a
complete $\TTOEW$ model with all the ingredients to explain the
diphoton hint observed by ATLAS and CMS. In Sec. \ref{sec:derivation}
we derive the low energy Lagrangian after $\TTOEW$ breaking, whereas
in Sec. \ref{sec:diphoton} we show how this setup naturally
accommodates the $X(750)$ state in a straightforward and natural way,
thus providing a complete framework for the diphoton anomaly. Finally,
we will conclude with a discussion.

\section{The model}
\label{sec:model}

We consider a variant of the models
in~\cite{Boucenna:2014ela,Boucenna:2015zwa}.  The model is based on
the $\TrTrOne$ gauge symmetry and contains three generations of lepton
triplets ($\psi_L$), two generations of quark triplets ($Q_L^{1,2}$),
one generation of quark anti-triplet ($Q_L^3$), plus their iso-singlet
right-handed partners. {The gauge symmetry breaking is achieved
through four scalar anti-triplets ($\Phi_{1,2,3}$ and $\Phi_{X}$).}
In our conventions, the electric charge generator is
defined as $Q=T_3+\frac{1}{\sqrt{3}}T_8+\mathcal{X}$ where $T_{3,8}$
are the diagonal generators of $SU(3)_L$.
\footnote{{That is, the $\TTOEW$ fundamental representations branch to $\EW$ as: \\
$(\mathbf{3}, \mathcal{X}) \to (\mathbf{2}, \tfrac{1}{6}+\mathcal{X}) \oplus (\mathbf{1},-\tfrac{1}{3}+\mathcal{X})$ 
and $(\overline{\mathbf{3}}, \mathcal{X}) \to (\mathbf{2},-\tfrac{1}{6}+\mathcal{X}) \oplus (\mathbf{1},\tfrac{1}{3}+\mathcal{X})$.}}
The particle content of the
model is summarized in~\tab{content}. \\

The fermions representations of the model can be decomposed as:
\begin{eqnarray} \label{eq:frep}
\psi_L &=& \left( \begin{array}{c}
\ell^- \\
- \nu \\
N^c \end{array} \right)_L^{e,\mu,\tau}\,,\nn\\
Q^1_L &=& \left( \begin{array}{c}
u \\
d \\
D \end{array} \right)_L\,,\,
Q^2_L = \left( \begin{array}{c}
c \\
s \\
S \end{array} \right)_L \,,\,
Q^3_L = \left( \begin{array}{c}
b \\
- t \\
T \end{array} \right)_L\,
\end{eqnarray}
The notation used for the extra quarks that constitute the third components of the $SU(3)_L$ triplets $Q^{1,2,3}_L$ is motivated by the fact that their electric charges are  $-1/3$ and $2/3$ for $D$/$S$ and $T$, respectively. The scalar multiplets can be written as:
\begin{equation} \label{eq:srep}
\Phi_1 = \left( \begin{array}{c}
\phi_1 \\
- \phi_1^+ \\
S_1^+ \end{array} \right) \,, \quad 
\Phi_{2,3} = \left( \begin{array}{c}
\phi_{2,3}^- \\
- \phi_{2,3} \\
S_{2,3}\end{array} \right) \,,  
\Phi_X = \left( \begin{array}{c}
\phi_X^- \\
- \phi_X \\
X \end{array} \right) \,.
\end{equation}
While $\phi_1^+$, $\phi_{2,3}^-$ and $S_1^+$ are electrically charged
scalars, the components $\phi_{1,2,3,X}$, $S_{2,3}$ and $X$ are
neutral. Therefore, neutral vacuum expectation values (VEVs) are
possible in the following directions,
$\vev{\Phi_1}^T=(k_1,0,0)/\sqrt{2}$,
$\vev{\Phi_2}^T=(0,k_2,n)/\sqrt{2}$,
$\vev{\Phi_3}^T=(0,k_3,n^\prime)/\sqrt{2}$, and
$\vev{\Phi_X}^T=(0,k_X,n_X)/\sqrt{2}$. However, in order to recover the
SM in the low energy limit, we assume the hierarchy $k_{1,2,3,X}\sim
v_{SM} \ll n, n^\prime, n_X$.  Moreover, we consider the particular
vacuum structure where $k_{2} = n^\prime = k_X = 0$ and $n_X =0$. The
first condition, together with a $\z2$ symmetry (see ~\tab{content}),
guarantees the existence of a simple pattern for quark masses, while
the second is required to explain the diphoton excess, as will be
clear below.~\footnote{
  Consistency with the minimization conditions of the scalar potential
  of the model has been checked explicitly.}
Therefore, the breaking of the gauge groups follows the chain
\begin{equation}
\TTOEW \xrightarrow{n}  \EW 
 \xrightarrow{k_{1,3}} U(1)_Q\,. \nonumber
\end{equation}

We note that while the lepton sector does not play an important role for what interests us here, it can nevertheless provide interesting complementary tests to probe the 3-3-1 scale via, for instance, neutrino masses and lepton flavor violation observables~\cite{Boucenna:2014ela,Boucenna:2015zwa}.

\begin{table}[!t]
\centering
\begin{tabular}{ c | c c c c c c c c  | c c c | c}
\toprule
& $\psi_L$ & $l_R$ & $Q_L^{1,2}$ & $Q_L^3$ & $q^u_R$ & $T_R$&$q^d_R$ &$D_R, S_R$  & $\Phi_1$ & $\Phi_2$ & $\Phi_3$&$\Phi_X$ \\ 
\hline
$SU(3)_C$ & $\one$ & $\one$ & $\three$ & $\three$ & $\three$ & $\three$ &$\three$&$\three$ & $\one$ & $\one$ & $\one$& $\one$\\
$SU(3)_L$ & $\threeS$ & $\one$ & $\three$ & $\threeS$ & $\one$ & $\one$ & $\one$ & $\one$  & $\threeS$ & $\threeS$ & $\threeS$ &$\threeS$\\
$U(1)_\mathcal{X}$ & $- \frac{1}{3}$ & $-1$ & $0$ & $+ \frac{1}{3}$ & $+ \frac{2}{3}$ & $+ \frac{2}{3}$ & $- \frac{1}{3}$ &$- \frac{1}{3}$ & $+ \frac{2}{3}$ & $- \frac{1}{3}$ & $- \frac{1}{3}$& $- \frac{1}{3}$\\[1mm]
\hline
$U(1)_{\mathcal{L}}$ & $- \frac{1}{3}$ & $-1$ & $- \frac{2}{3}$ & $+ \frac{2}{3}$ & $0$ & $0$ & $0$& $0$ & $+ \frac{2}{3}$ & $- \frac{4}{3}$ & $+ \frac{2}{3}$  & $- \frac{4}{3}$ \\
$\z2$ & $+$ & $+$ & $+$  & $-$ & $+$&$-$ & $-$ & $+$& $+$ &$+$&$-$&$+$ \\
\bottomrule
\end{tabular}
\caption{Particle content of the model. Here $q^u_R\equiv \{u_R, c_R,
  t_R\}$, $q^d_R\equiv \{d_R,s_R,b_R\}$. The global symmetry
  $U(1)_{\mathcal{L}}$ allows for a consistent definition of lepton
  number via the relation $L=\frac{4}{\sqrt{3}} T_8+\mathcal{L}$ and
  the $\z2$ parity simplifies the expressions of quark masses. See
  Ref.~\cite{Boucenna:2014ela} for further details.}
\label{tab:content}
\end{table}

\section{Phenomenological Lagrangian from $\TTOEW$}
\label{sec:derivation}

It is instructive to write the Lagrangian in the $\EW$ symmetric
phase, i.e., after $\TTOEW$ gets broken at a high-energy scale
$\vev{S_2}$. Before symmetry breaking, the quark Lagrangian
reads:
\begin{eqnarray}
\mathcal{L}_{\text{quarks}} &=&
\bar Q_L^{1,2}\, y^u  q^u_R \Phi_1^\ast + \bar Q_L^{3}\, \tilde y^d q^d_R \Phi_1+ + \bar Q_L^{1,2}\, \bar y^d   \hat d_R \Phi_2^\ast \nn\\
&&+  \bar Q_L^3\, \bar y^u  T_R \Phi_2 +\,\bar Q_L^3\, \tilde y^u  q^u_R \Phi_3 + \bar Q_L^{1,2}\, y^d q^d_R \Phi_3^\ast  \nn\\
&&+ \bar Q_L^{1,2}\, \bar y^d   \hat d_R \Phi_X+ \bar Q_L^3\, \bar y^u  T_R \Phi_X   + 
\, \hc\,,
\label{eq:lagqrk}
\end{eqnarray}
where we defined $\hat d_R \equiv ( D_R, S_R)$, $q^u_R\equiv \{u_R,
c_R, t_R\}$ and $q^d_R\equiv \{d_R,s_R,b_R\}$. After $\TTOEW$ gets
broken to $\EW$, the $SU(3)_L$ triplets split into doublet and singlet
representations of $SU(2)_L$. {We can write} \eq{lagqrk} as:
\begin{equation}
\label{eq:Lqtot}
\mathcal{L}_{\text{quarks}} = \mathcal{L}_{\text{SM-quarks}} + \mathcal{L}_{S_1^+} + \mathcal{L}_{S_2}+ \mathcal{L}_{S_3} + \mathcal{L}_{X} \,,
\end{equation}
\begin{figure}[t!]
\label{fig:cmq}
\begin{center}
\includegraphics[width=.9\linewidth]{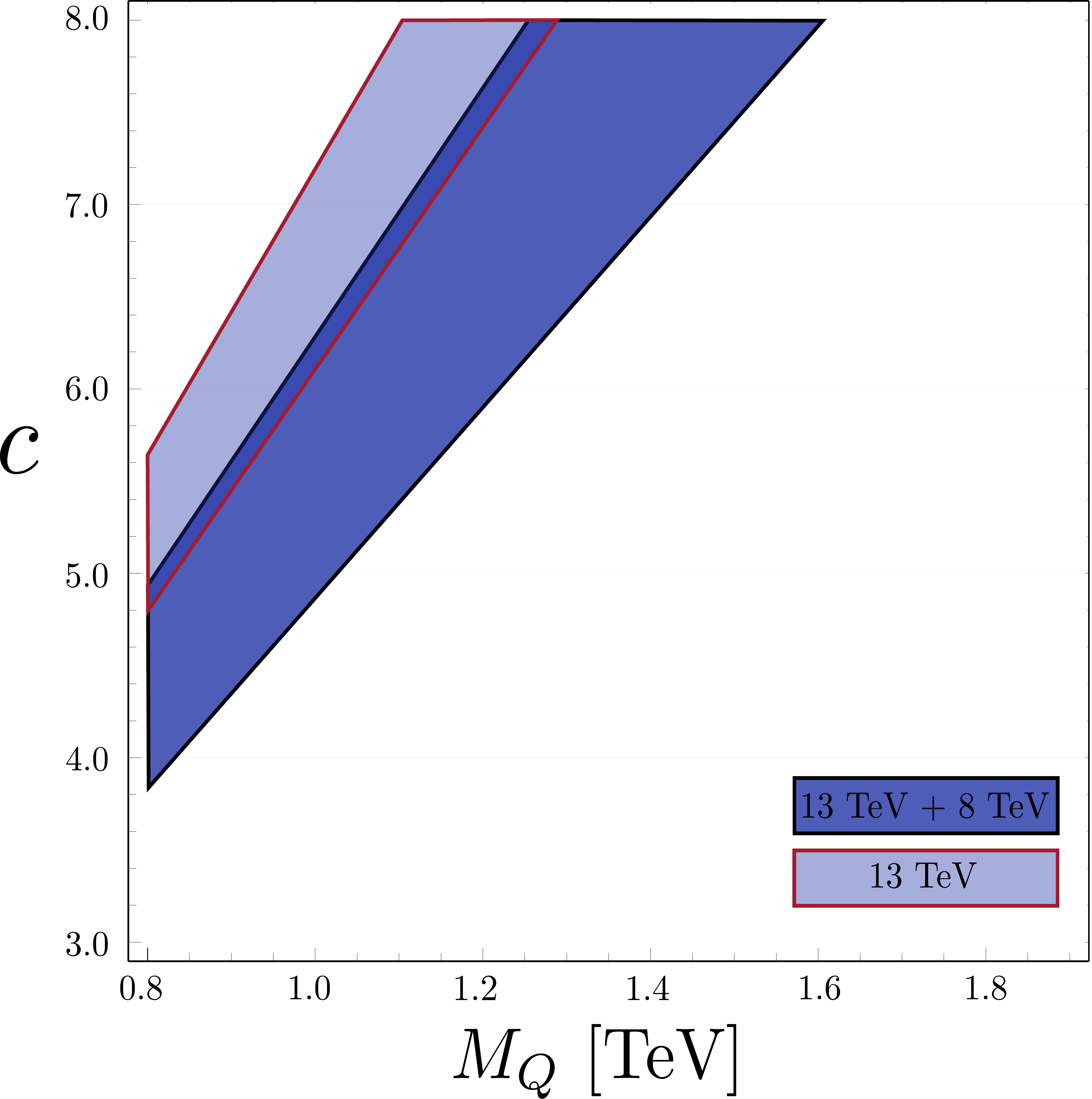}
\caption{The coupling between $X$ and the VL quarks, $c$, as a
  function of VL quarks mass $M_Q$. All the points satisfy the bound
  $M_Q>800 \gev$ and exclusion limits on $\sigma(gg\to X) \times
  \text{BR}(X\to VV)$. The bands correspond to the $95\%$ regions of
  the combined ATLAS and CMS data for $\sigma(gg\to X) \times
  \text{BR}(X\to \gamma\gamma)$ with 13 TeV ($(6.1 \pm 1)$ fb
  \cite{Ellis:2015oso}, dark) and 13 TeV + 8 TeV ($(4.4 \pm 1.1)$ fb
  \cite{Buttazzo:2015txu}, light). }
\label{fig:1}
 \end{center}
\end{figure}
where we denote all the terms involving SM quarks and $SU(2)_L$
Higgses (scalar doublets) by $\mathcal{L}_{\text{SM-quarks}}$. This
includes for instance terms such as $\overline{q_L^{1,2}}\,i \sigma_2\, H_2^\ast\,
\hat d_R$, with $\sigma_2$ the second Pauli matrix and $H_2^\tr \equiv (\phi_2\quad \phi_2^-)$. The terms in
$\mathcal{L}_{S_1^+}$ involve a charged scalar and so we ignore them
here. The terms involving interactions with the neutral singlet
$S_3$ are:
\begin{equation}
\label{eq:LXp}
\mathcal{L}_{S_3} =
c_d'' \,  S_3^\ast \, \overline{D}_L\,  q^d_R
+
 c_s'' \, S_3^\ast \, \overline{ S}_L\,   q^d_R
+
 c_t'' \,  S_3 \, \overline{T}_L\,  \,   q^u_R
+\, \hc\,,
\end{equation}
where we have simplified the notation of the couplings.
Since $\vev{S_3}=0$, the $S_3$ particle
cannot be produced via gluon fusion to quark loops (i.e., there is no
mixing between SM quarks and the new ones~\footnote{We notice that our
  choice for the vacuum structure of the model guarantees that the
  exotic quarks that constitute the third components of the
  $SU(3)_L$ triplets do not mix with the SM states after symmetry
  breaking. This can be seen in \app{quarks}, where the quark mass
  matrices are explicitly derived.} nor can it decay to photons and
thus cannot account for the $X(750)$ resonance).  We are then left with
two candidates for $X(750)$: $X$ and $S_2$, giving the low-energy
Lagrangians after $\TTOEW$ breaking
\begin{eqnarray}
\label{eq:LX}
\mathcal{L}_{S_2}&=& 
c_d^\prime \,  S_2^\ast \, \overline{D}_L\,  D_R
+
 c_s^\prime \, S_2^\ast \, \overline{ S}_L\,   S_R
+
 c_t^\prime \,  S_2 \, \overline{T}_L\,  T_R
+\, \hc\,,\nn \\
\\ 
\mathcal{L}_{X}&=& 
c_d \,  X^{\ast} \, \overline{D}_L\,  D_R
+
 c_s \, X^{\ast} \, \overline{ S}_L\,   S_R
+
 c_t \,  X \, \overline{T}_L\,  T_R
+\, \hc\nn\,\\
\end{eqnarray}
Again, $c_q^{(\prime)}$ are a simplified notation of the components of
the coupling matrices appearing in \eq{lagqrk}, which we take to be
diagonal for simplicity. For instance we defined $\bar
y^{d}_{14}\equiv c_d$, $\bar y^{d}_{25}\equiv c_s$, $\bar
y^{u}_{14}\equiv c_t$, and $y^{d}_{15}=y^{d}_{24}=0$. From these two
possibilities only $X$ is able to reproduce the $X(750)$ resonance as
$S_2$ would imply an unacceptably low $\TTOEW$ breaking scale,
see \secs{diphoton}. 

We refer to \app{scalar} for a detailed discussion of the scalar and
gauge boson spectra.  It is clear then that $\mathcal{L}_{X}$ can be
seen as an effective Lagrangian (or `simplified model') extending the
SM with a { neutral iso-singlet} scalar, and pairs of vector-like quarks transforming as
$(\three,\one)_{-1/3}$ for $D$ and $S$ and as $(\three,\one)_{2/3}$
for $T$.  $X$ can be produced through gluon fusion and decay to
photons via triangle loops involving these new quarks and is therefore
our natural candidate for the diphoton resonance. Variants of this
effective Lagrangian have been analyzed, see for example
\cite{Buttazzo:2015txu,Ellis:2015oso,Falkowski:2015swt}, and have been
shown to be able to account for the diphoton excess. In the next
section we derive these results for our specific model.

\section{The diphoton excess}
\label{sec:diphoton}
The $X$ particle can decay to a pair of quarks or gauge bosons. {Other
  decay channels are either kinematically inaccessible or have very
  suppressed widths, as explained below.} Taking all the masses of the
new quarks to be $M_Q$, we can approximately express the widths to
gluons and gammas as: \cite{Buttazzo:2015txu}
\begin{eqnarray}
\label{eq:widths}
\Gamma(X\to g g) &\simeq& \frac{2 M_X \alpha_s^2}{9 \pi} \left(\frac{M_X}{4\pi M_Q}\right)^2  \left|\sum_{i=d,s,t}{c_i}\right|^2 \nn\\
&=& \frac{4 M_X \alpha_s^2}{\pi} \left(\frac{M_X}{4\pi n}\right)^2\, \left(\frac{c}{c^\prime}\right)^2\,,\\  
\Gamma(X\to \gamma \gamma) &\simeq&   \frac{M_X  \alpha_e^2 }{\pi} \left(\frac{M_X}{4\pi M_Q}\right)^2\left|\sum_{i=d,s,t}{c_i Q_i^2}\right|^2\nn\\
&=&\frac{4 M_X  \alpha_e^2 }{3 \pi} \left(\frac{M_X}{4\pi n}\right)^2 \, \left(\frac{c}{c^\prime}\right)^2\,,
\end{eqnarray}
The second equality follows from the assumption of universal couplings $c_i =c$, $c_i^\prime=c^\prime$,
and the relation $M_Q=c^\prime n/\sqrt{2}$.
It follows that:
\begin{eqnarray}
\label{eq:prop}
\Gamma(X\to \gamma\gamma) &=& \frac{1}{3} \left(\frac{\alpha_e}{\alpha_s}\right)^2 \, \Gamma(X\to gg) \\
&\approx& 2.1\times 10^{-3}  \, \Gamma(X\to g g)\nn\,,
\end{eqnarray}
where we used $\sqrt{4 \pi \alpha_s} = 1.07$ at the energy scale $Q =
750$ GeV and $\sqrt{4 \pi \alpha_e}= 0.30$ at $Q = 0$.
Therefore, in the absence of other decay channels, the branching ratio
to photons is $\text{BR}_{\gamma \gamma} \approx 2.1 \times 10^{-3}$.
The decay widths into $Z\gamma$ and $ZZ$ via quark loops are always
smaller than the $\gamma\gamma$ decay width because they proceed via
electroweak mixing; for $SU(2)_L$-singlet VL fermions the branching
ratios are fixed to be: $\text{BR}_{\gamma \gamma} : \text{BR}_{\gamma
  Z}:\text{BR}_{Z Z} = 1: 2 \tan_W^2: \tan_W^4$, with $\tan_W \approx
0.55$. These rates are in agreement with ATLAS and CMS bounds.
The tree-level decay to $ZZ$ is induced by $Z-Z'$ mixing ($\propto
(k/n)^2$), and hence suppressed by the 3-3-1 breaking scale. We found
it to be small unless $n\lesssim 1.3 \tev$, which would be in conflict
with bounds on $Z'$ direct searches at LHC which are in the multi-TeV
range~\cite{Martinez:2014lta,Salazar:2015gxa}.  This also excludes the
possibility of a significant contribution of gauge bosons loops to the
diphoton signal.
Decays to $WW$ are not present because there is no $W-W'$ mixing due to the
underlying gauge symmetry.\\

\begin{figure}[t!]
\label{fig:ccp}
\begin{center}
\includegraphics[width=.9\linewidth]{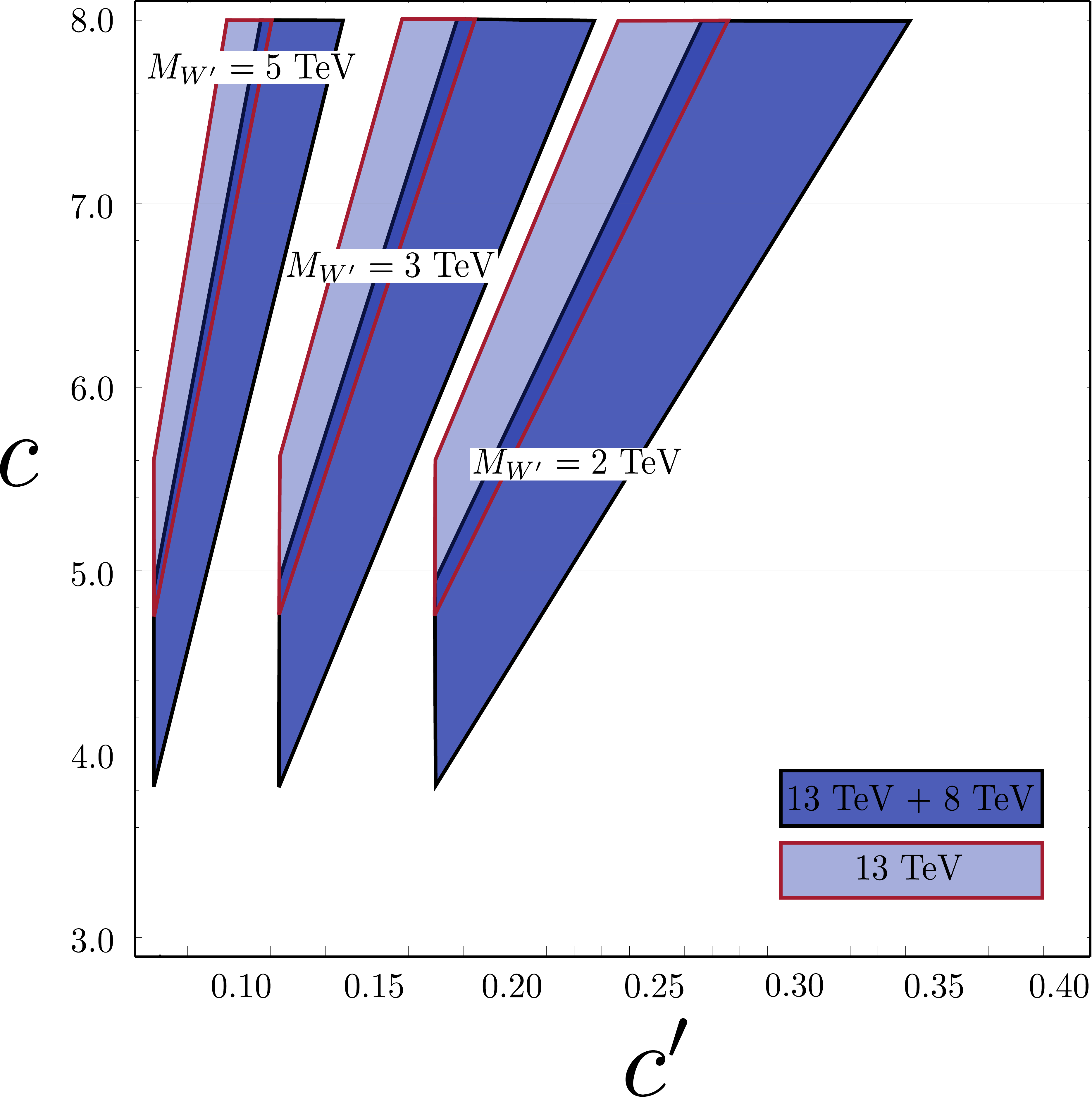}
\caption{The coupling between $X$ and the VL quarks, $c$, as a
  function of the coupling $c^\prime$ ($M_Q=c^\prime n/\sqrt{2}$) for
  different masses of the $W^\prime$ gauge boson. All the points
  satisfy the bound $M_Q>800 \gev$ and exclusion limits on
  $\sigma(gg\to X) \times \text{BR}(X\to VV)$. The bands correspond to
  the $95\%$ regions of the combined ATLAS and CMS data for
  $\sigma(gg\to X) \times \text{BR}(X\to \gamma\gamma)$ with 13 TeV
  ($(6.1 \pm 1)$ fb \cite{Ellis:2015oso}, dark) and 13 TeV + 8 TeV
  ($(4.4 \pm 1.1)$ fb \cite{Buttazzo:2015txu}, light). }
\label{fig:2}
 \end{center}
\end{figure}

In fig. (\ref{fig:1}) we show the variation of the coupling $c$ as a
function of the VL quark mass $M_Q$ that satisfies the data. We take
the $95\%$ C.L. regions on the combined ATLAS and CMS data using only
13 TeV data from Run II \cite{Ellis:2015oso} or a combination of 13
TeV and 8 TeV data \cite{Buttazzo:2015txu}.  On the other hand, in
fig. (\ref{fig:2}) we show for various $M_{W^\prime}$ the required
coupling $c^\prime$ in order to fit the data. The lower-bound on
$c^\prime$ translates the bound $M_Q>800 \gev$ on VL quarks. For these
figures we have used the exact leading-order relation for $\Gamma(X\to
\gamma \gamma)$ (see \app{gamma}) instead of the approximation in
\eq{widths} since $M_Q \sim M_X$. Furthermore, we estimate the
production cross-section $\sigma(gg\to X)$ adapting the results of
\cite{Buttazzo:2015txu}, and we have explicitly checked that these are
compatible with those in \cite{Franceschini:2015kwy,Ellis:2015oso}. We
see that the coupling of $X$ with the new quarks has to be relatively
large, $\sim 5$ for $M_Q\sim 0.8\tev$, in order to accommodate the 13
TeV data ($\sim 4$ if one considers the combined 13 TeV and 8 TeV
data). Still, compared to the case where only one down(up)-type quark
is present, the improvement is significant since that would have
required couplings as large as $\sim 35 (9)$. Also, we note that the
physical $X \overline{\mathcal{Q}} \mathcal{Q}$ vertex is
$c/\sqrt{2}$, and not just $c$, in the perturbative regime.

Another result that can be extracted from fig. (\ref{fig:2}) is that a
hierarchy between $c$ and $c^\prime$ is required in order to explain
the diphoton hint. This excludes what could be seen as the minimal
possibility of our framework to explain $X(750)$, namely
$S_2$. Indeed, in a simpler model without the $\Phi_X$ triplet,
$c = c^\prime$ and {the dependence on the coupling is
very weak in the decay widths, as can be seen in the approximate
relations \eq{widths} where it completely disappears}. {This means that
if the signal is interpreted as arising from the decay of a
scalar via VL quarks loops, then this scalar cannot be the origin of
the VL quarks masses in the context of 3-3-1 models with 
non-exotic quark charges}.

Finally, we comment on the width of the $X(750)$ resonance. Although
the ATLAS fit seems to improve if the width is large, $\Gamma \sim 6\%
M_X$, current data are perfectly compatible with the existence of a
narrow resonance. Nevertheless, we have investigated whether the model
could simultaneously account for a large width and the diphoton
signal. In principle, the extra decay channel could be provided via
the leptonic term $\mathcal{L}_{\text{leptons}} \supset y^s\, X
\overline{N_L^c} N_S $~\cite{Boucenna:2015zwa}, which is required to
generate neutrino masses via the inverse-seesaw
mechanism~\cite{Mohapatra:1986bd}.\footnote{Here $N_S$ denotes a Majorana
$SU(3)_L$ singlet (whose majorana mass is the usual `$\mu$ term' of
the inverse-seesaw).} However, we have found that such term cannot
increase the width of the $X(750)$ scalar so as to reach the best-fit
value found by ATLAS. Therefore, if future data shows a clear
preference for a broad resonance, an extension of our setup will
be required.

\section{Discussion and conclusions}
\label{sec:conclusion}
In this article we have shown that a simple gauge extension of the SM
can account for the diphoton excess recently observed by ATLAS and
CMS. The gauge structure of the theory requires $3$ extra quarks to
complete the fundamental $SU(3)_L$ multiplets. These quarks are
effectively $SU(2)_L$ singlet VL quarks at low energies. If coupled to
a fundamental scalar (anti-)triplet that does not contribute
dominantly to their mass (if it takes VEV) then the low-energy
iso-singlet explains naturally the $X(750)$ resonance.\\

The multiplicity of the new quarks, due to the number of families,
reduces the severe requirement on the coupling between $X$ and the
quarks. This translates as a possible perturbative explanation for
the diphoton anomaly.
{ Moreover, we find that $X$ cannot be responsible of the diphoton signal and the breaking
of $\TTOEW$ to the electroweak gauge group at the same time, as that would be excluded by
$Z'$ direct searches.}\\

To conclude, we emphasize once again that more data is required in
order to fully assess the relevance of the diphoton excess. If
confirmed, exciting times will come 
{ in the quest of New Physics} that expands our current understanding of particles physics. Indeed, $X$
may well be the tip of the iceberg, and future data of LHC will reveal
other particles from the UV completion of the theory, possibly in the
form of new colored particles and new gauge bosons which are all lying
around the TeV scale in our framework.

\section*{Note added}

Shortly after the appearance of our paper, other explanations for the
diphoton excess based on the $\TTOEW$ gauge symmetry were proposed in
\cite{Cao:2015scs,Hernandez:2015ywg,Dong:2015dxw}. In contrast to the
specific model introduced here, these references consider 3-3-1 models
with exotic electric charges, typically leading to slightly lower
Yukawa couplings to explain the diphoton excess.

\section*{Acknowledgments}
S.M.B acknowledge support of the MIUR grant for the Research
Projects of National Interest PRIN 2012 No. 2012CP- PYP7 {\it
  Astroparticle Physics} and of INFN I.S. {\it TASP}
2014 and support of the Spanish MICINN's
Consolider-Ingenio 2010 Programme under grant MultiDark
CSD2009-00064. A.V. acknowledgees financial support by the Spanish
grants FPA2014-58183-P, Multidark CSD2009-00064 and SEV-2014-0398
(MINECO), FPA2011-22975 and PROMETEOII/2014/084 (Generalitat
Valenciana) and is grateful to Martin Hirsch, Diego Aristiz\'abal
Sierra and Juan Herrero-Garc\'ia for enlightening discussions about
the diphoton excess and to Isabel Cordero-Carri\'on for her amazing support.

\appendix

\section{$X\to \gamma \gamma$ width}
\label{app:gamma}

Following \cite{Spira:1995rr,Carena:2012xa}, the diphoton decay width
of the scalar $X$ to two photons via a loop with particles $D$, $S$,
$T$, all with mass $M_Q>M_X/2$, is given by:
\begin{equation}
 \label{eq:exact}
  \Gamma(X\to \gamma\gamma) = \frac{\alpha_e M_X^3}{512 \,\pi^3 \, M_Q^2} \left| 
 N_c\, \sum_{i=D,S,T}{c_i} Q_i^2 A_{1/2} (\tau_i)
  \right|^2\,,
\end{equation}
with $\tau_i = 4 M_{Q_i}^2/M_X^2$, $N_c=3$, $Q_i$ the electric charges
of the heavy quarks and
\begin{eqnarray}
 A_{1/2} (\tau) &=& 2 \tau^2 \left( \tau^{-1} + \left(\tau^{-1}-1 \right) f(\tau^{-1}) \right) \, , \\
 f(x) &=& \arcsin^2 \sqrt{x} \, .
\end{eqnarray}

We note that the last expression is valid for a $X(750)$ scalar whose
mass is below the kinematic threshold for the production of two heavy
$Q$ states.

\section{Scalar potential and mass spectrum}
\label{app:scalar}

The scalar potential of the model can be written as:~
\footnote{For the sake of clarity and simplicity, we consider all the terms involving a single power of $\Phi_{2,X}$ as either absent or with a very small coupling, which is technicaly natural since this enhances the symmetries of the potential. We also assume CP conservation in the scalar potential.}
\begin{eqnarray}
\label{eq:V}
V  &=& \sum_i \mu_i^2 |\Phi_i|^2 + \lambda_i |\Phi_i|^4 + \sum_{i \ne j} \lambda_{ij} |\Phi_i|^2|\Phi_j|^2 \nonumber\\
&& + f\, (\Phi_1 \Phi_2 \Phi_3+\hc)  + \frac{\kappa}{2} \left[ (\Phi_2^\dagger \Phi_X)^2 +\hc \right] \, , \nn\\
\end{eqnarray}
where $i=1,2,3,X$.  The $\z2$-soft-breaking term, $f \Phi_1 \Phi_2
\Phi_3$, is required to break accidental symmetries appearing in the
scalar potential. 
{ Since $\Phi_2$ and $\Phi_X$ have identical quantum numbers, all the operators in the scalar potential
remain invariant in the exchange $\Phi_2 \leftrightarrow \Phi_X$, however we have assumed in eq.~(\ref{eq:V}) that operators
involving only one power of $\Phi_X$ are absent (either because their couplings are so small that they can be ignored, or because
of a symmetry which distinguishes $\Phi_2$ from $\Phi_X$, e.g. a parity) in order to avoid tree-level mixing between $X$ and the SM Higgses
and allow for the particular vacuum solution with $n_X = 0$.}
Assuming $k_1 \sim k_3 \equiv k \ll n$ and universal couplings, we
find that the mass spectrum of the charged scalars is given at leading
order as: \footnote{We identify the corresponding (approximate)
  eigenstates between parentheses.}
\begin{eqnarray}
M^2(\phi_2^\pm) &=&0 \,,\\
M^2((\phi_1^\pm + \phi_3^\pm)/\sqrt{2})&=&0 \,,\\
M^2((\phi_1^\pm - \phi_3^\pm)/\sqrt{2})&\sim& \frac{1}{\sqrt{2}}f\, n \,,\\ 
M^2(S_1^\pm) &\sim& \sqrt{2} f\, n \,, \\
M^2(\phi_X^\pm) &=& \mu_X^2 + \lambda \, k^2 \,.
\end{eqnarray}
The masses of the neutral CP-even scalars are, up to corrections of $\order{k^2}$:
\begin{eqnarray}
M^2(\Re(\phi_1 + \phi_3)/\sqrt{2})&\sim&(2 \lambda + \sqrt{2}\, \frac{f}{n} - \frac{1}{2}\,\frac{f^2}{\lambda n^2} ) k^2\,,\nn \\
\\
M^2(\Re(\phi_1 - \phi_3)/\sqrt{2})&\sim&  \frac{1}{\sqrt{2}} f \, n\,,\\
M^2(\Re\phi_2)&=&0    \,,\\
M^2(\Re{S_2}) &\sim&2 \lambda\, n^2 \,,\\
M^2(\Re{S_3}) &\sim& \sqrt{2} f \, n  \,,\\
M^2(\Re\phi_X)&=& \mu_X^2 + \lambda \, k^2    \,.
\end{eqnarray}
On the other hand, the masses of the neutral CP-odd scalars at leading order are
given as:
\begin{eqnarray}
M^2(\Im(\phi_1 + \phi_3)/\sqrt{2})&\sim& \sqrt{2} f\,n \,,\\
M^2(\Im(\phi_1 - \phi_3)/\sqrt{2})&=& 0 \,,\\
M^2(\Im\phi_2) &=&0 \,,\\
M^2(\Im{S_2}) &=&0 \,,\\
M^2(\Im{S_3}) &\sim& \frac{1}{\sqrt{2}} f \, n  \,, \\
M^2(\Im\phi_X)&=& \mu_X^2 + \lambda \, k^2    \,.
\end{eqnarray}
Finally, the state $\Re{X} \equiv X$ has a mass:
\begin{align}
M^2(\Re{X}) &= M_X^2 \\
&= \mu_X^2 + \frac{1}{2} \left( \lambda_{1X} + \lambda_{3X}\right) k^2 + \frac{1}{2} \kappa \, n^2 \nonumber
\end{align}
and does not mix with the other CP-even scalars. The mass of its
CP-odd counterpart, $\Im{X}$, can be independently set with a proper
choice of the $\kappa$ quartic coupling, as one finds $M^2(\Re{X}) -
M^2(\Im{X}) = \kappa n^2$. The massless scalars found in the above
equations correspond to the degrees of freedom `eaten-up' by the
charged and neutral gauge bosons, respectively, which acquire the
following masses:
\begin{eqnarray}
m_{W}^2  &=& \frac{1}{2} \, g_2^2 \, k^2 \,, \\
m_{W'}^2  &=& \frac{1}{4} \, g_2^2 \, n^2 \,,\\
m_Z^2\,\,    &=&  \frac{g_2^2(4 g_1^2+3 g_2^2)}{2 \left(g_1^2+3 g_2^2 \right)}\,k^2  \,, \\
m_{Z'}^2 &=& \frac{1}{9} \, (g_1^2+ 3 g_2^2) \, n^2 \,, \\
m_{X_G}^2  &=& m_{Y_G}^2 \, = \, \frac{1}{4} \, g_2^2 \, n^2\,.
\end{eqnarray}
{with $g_2 (g)$ being the coupling constant of $SU(3)_L$ ($U(1)_\mathcal{X}$)}.
Notice that since $S_2$ is singlet under the $SU(2)_L$
subgroup contained in $SU(3)_L$, the VEV $n$ { will control the four
  new gauge bosons masses} and break $SU(3)_L$ to $SU(2)_L$.  On the
other hand, $\EW$ is broken at the electroweak scale by the $k_1$ and
$k_3$ VEVs down to the electromagnetic $U(1)_Q$ symmetry.
For $f\sim n$ all the scalars of the model are naturally
heavy, except one state that we can identify with the SM Higgs boson,
i.e., $H\equiv (\phi_1 + \phi_3)/\sqrt{2}$, in good
approximation. Indeed, its couplings to the fermions confirm that the
state $H$ is the one that gives mass to the SM fermions.

\section{Quark masses}
\label{app:quarks}

The quark Lagrangian in \eq{lagqrk} leads to the following mass matrices:
\begin{equation}
M_d = - \frac{1}{\sqrt{2}} \left( \begin{array}{ccccc}
y^d_{11} \, k_2 & y^d_{12} \, k_2 & y^d_{13} \, k_2 & 0 & 0 \\
y^d_{21} \, k_2 & y^d_{22} \, k_2 & y^d_{23} \, k_2 & 0 & 0 \\
\tilde y^d_{11} \, k_1 & \tilde y^d_{12} \, k_1 & \tilde y^d_{13} \, k_1 & 0 & 0 \\
0 & 0 & 0 & \bar y^d_{14} \, n & \bar y^d_{15} \, n \\
0 & 0 & 0 & \bar y^d_{24} \, n & \bar y^d_{25} \, n
\end{array} \right)\,,
\end{equation}
\begin{equation}
M_u = - \frac{1}{\sqrt{2}} \left( \begin{array}{cccc}
y^u_{11} \, k_1 & y^u_{12} \, k_1 & y^u_{13} \, k_1 & 0 \\
y^u_{21} \, k_1 & y^u_{22} \, k_1 & y^u_{23} \, k_1 & 0 \\
\tilde y^u_{11} \, k_2 & \tilde y^u_{12} \, k_2 & \tilde y^u_{13} \, k_2 & 0 \\
0 & 0 & 0 & \bar y^u_{14} \, n
\end{array} \right) \, .
\end{equation}
The $\z2$ symmetry, see \tab{content}, is introduced so that the SM
quarks and the new ones are independent of each other and can be
adjusted individually to easily obtain a realistic quark sector and
heavy exotic quarks at the same time.

\bibliography{refs}

\end{document}